\documentclass[11pt,a4paper]{article}  
\usepackage{amsmath}
\usepackage{amsfonts,amssymb}
\begin{document}

\newcommand{\nl}{\nonumber\\}
\newcommand{\nnl}{\nl[6mm]}
\newcommand{\nle}{\nl[-2.3mm]\\[-2.3mm]}
\newcommand{\nlb}[1]{\nl[-2.3mm]\label{#1}\\[-2.3mm]}

\renewcommand{\theequation}{\thesection.\arabic{equation}}
\let\ssection=\section
\renewcommand{\section}{\setcounter{equation}{0}\ssection}

\newcommand{\be}{\bes}
\newcommand{\ee}{\ees}
\newcommand{\bes}{\begin{eqnarray}}
\newcommand{\ees}{\end{eqnarray}}
\newcommand{\eens}{\nonumber\end{eqnarray}}

\renewcommand{\/}{\over}
\renewcommand{\d}{\partial}

\newcommand{\eps}{\epsilon}
\newcommand{\vth}{\vartheta}
\renewcommand{\th}{\theta}
\newcommand{\dlt}{\delta}
\newcommand{\al}{\alpha}
\newcommand{\si}{\sigma}
\newcommand{\w}{\omega}
\newcommand{\ww}{\varpi}

\newcommand{\bx}{\bar x}
\newcommand{\bd}{\bar\d}

\newcommand{\tD}{\widetilde D}
\newcommand{\tE}{\widetilde E}

\renewcommand{\L}{{\cal L}}

\newcommand{\oj}{{\mathfrak g}}
\newcommand{\tr}{\hbox{tr}\,}
\newcommand{\rank}{\hbox{rank}\,}
\newcommand{\inv}[1]{\hbox{$1\/#1$}}
\newcommand{\half}{\hbox{$1\/2$}}
\newcommand{\third}{\hbox{$1\/3$}}
\newcommand{\fourth}{\hbox{$1\/4$}}
\newcommand{\fifth}{\hbox{$1\/5$}}
\newcommand{\sixth}{\hbox{$1\/6$}}
\newcommand{\eighth}{\hbox{$1\/8$}}

\newcommand{\ssg}{sl(3)\oplus sl(2)\oplus gl(1)}

\newcommand{\vect}{{\mathfrak{vect}}}
\newcommand{\svect}{{\mathfrak{svect}}}
\newcommand{\hh}{{\mathfrak{h}}}
\newcommand{\kk}{{\mathfrak{k}}}

\newcommand{\ggl}{{\mathfrak{gl}}}
\newcommand{\ssl}{{\mathfrak{sl}}}
\newcommand{\ssp}{{\mathfrak{sp}}}
\newcommand{\sso}{{\mathfrak{so}}}
\newcommand{\poin}{{\mathfrak{poin}}}
\newcommand{\co}{{\mathfrak{co}}}
\newcommand{\e}{{\mathfrak{e}}}
\newcommand{\f}{{\mathfrak{f}}}
\newcommand{\g}{{\mathfrak{g}}}

\newcommand{\mb}{{\mathfrak{mb}}}
\newcommand{\sle}{{\mathfrak{sle}}}
\newcommand{\ksle}{{\mathfrak{ksle}}}
\newcommand{\vle}{{\mathfrak{vle}}}
\newcommand{\kas}{{\mathfrak{kas}}}
\newcommand{\vas}{{\mathfrak{vas}}}

\newcommand{\barr}{\begin{array}}
\newcommand{\earr}{\end{array}}

\newcommand{\CC}{{\mathbb C}}
\newcommand{\ZZ}{{\mathbb Z}}
\newcommand{\GG}{{\mathcal G}}

\title{{Structures Preserved by Exceptional Lie Algebras}}

\author{T. A. Larsson \\
Vanadisv\"agen 29, S-113 23 Stockholm, Sweden,\\
 email: Thomas.Larsson@hdd.se }

\maketitle 
\begin{abstract} 
For $\ssp(n+2)$ and each exceptional Lie algebra a realization of depth $2$
preserving the spaces spanned by a contact one-form and a bilinear form is
given. For $\e_7$ and $\e_6$ a realization of depth $1$ preserving a
lightcone and the space spanned by a bilinear form is also presented. This
makes the origin of the exceptions clear.
\end{abstract}

\renewcommand{\arraystretch}{1.4}

\section{Introduction}

The exceptional simple Lie algebras $\e_8$, $\e_7$, $\e_6$, $\f_4$ and
$\g_2$ were discovered over a century ago. Although many things are
known about them, they still remain somewhat of a mystery. The purpose
of this work is to describe them in a way that makes their
existence obvious and inevitable.

In \cite{GKN00} the exceptions were presented as algebras of vector
fields, i.e. subalgebras of $\vect(n)$ for some $n$. $\e_7$ and $\e_6$
can be described as a graded Lie algebra of depth $1$ (three-graded
structure, conformal realization):
\[
\oj = \oj_{-1} + \oj_0 + \oj_1,
\]
and every exception can be described as a graded Lie algebra of
depth $2$ (five-graded structure, quasiconformal 
realization):\footnote{Unfortunately, standard notation for the 
degree $2$ subspace coincides with the exception $\g_2$.}
\[
\oj = \oj_{-2} + \oj_{-1} + \oj_0 + \oj_1 + \oj_2.
\]
Unfortunately, vector fields of positive degree are non-linearly 
realized, which makes the description in \cite{GKN00}
quite opaque. However, it is known that the classical simple Lie 
algebras, both the finite- and infinite-dimensional ones, are best
described as vector fields preserving some structure, be it a
differential form, a bilinear form, or a subspace spanned by certain 
forms. In the 
subsequent sections similar descriptions for the exceptions are given.

The main idea is that the preserved structures are completely
determined by vector fields in $\oj_{-2} + \oj_{-1} + \oj_0$.
This is the basis for the method of Cartan
prolongation, extensively developed by Leites and Shchepochkina in the
context of infinite-dimensional Lie superalgebras \cite{ALS80,Sh99}.
The strategy used in the present paper can thus be summarized as follows:
\begin{enumerate}
\item
Find a realization of $\oj_{-2} + \oj_{-1} + \oj_0$ on $\CC^n$; in most
cases, this has already been done in \cite{GKN00}.
\item
Determine the structures preserved by these vector fields.
\item
Define $\oj$ as the subalgebra of $\vect(n)$ preserving
the same structures.
\end{enumerate}
This strategy was employed to find realizations of three exceptional
Lie superalgebras in \cite{Lar01}.

This method has the drawback that one can not be completely certain that
the resulting algebra is the right one. However, vector fields
preserving some structure automatically define a closed subalgebra, and
this subalgebra is not empty since it contains the right vector fields
at non-positive degree. Moreover, and unlike the algebras themselves,
the preserved structures have very simple and natural descriptions,
which makes me believe that this is the best way to understand the
exceptional Lie algebras.

All infinite-dimensional simple Lie algebras of vector fields can also
be described as graded algebras, i.e.
\[
\oj = \oj_{-d} + ... + \oj_{-1} + \oj_0 + \oj_1 + \oj_2 + ...,
\]
where the depth $d \leq 2$; there are two Lie superalgebras of depth $3$.
Regarding simple Lie algebras as algebra of vector fields thus gives a
unified description of the finite- and infinite-dimensional cases.
In fact, the distinction between finite and infinite dimension seems
somewhat artificial. The Poincar\'e and Hamiltonian algebras are
examples of the two types, and yet they preserve very similar 
structures; the square length element and the symplectic two-form only
differ in their symmetries.

All results in this paper are formulated on the Lie algebra level, but
they should readily generalize to the corresponding groups $E_8$, $E_7$,
$E_6$, $F_4$ and $G_2$. Namely, if $\oj \subset \vect(n)$ preserves some
structure, $G$ should be the subgroup of the diffeomorphism group
$Diff(\CC^n)$ that preserves the same structure. Moreover, I expect that
the results can be globalized, i.e. that manifolds with exceptional 
structures exist in certain dimensions, just like symplectic and contact
manifolds correspond to algebras of Hamiltonian and contact vector fields.

Tensor calculus notation is used throughout this paper. Repeated
indices, one up and one down, are implicitly summed over. Symmetrization
is denoted by parentheses, $a^{(i}b^{j)} \equiv a^ib^j + a^jb^i$ and
anti-symmetrization by brackets, $a^{[i}b^{j]} \equiv a^ib^j - a^jb^i$.
The Kronecker delta $\dlt^i_j$ and the totally anti-symmetric constant
$\eps^{i_1i_2..i_n}$ with inverse $\eps_{i_1i_2..i_n}$ are the only
invariant tensors under $\vect(n)$.

The base field is $\CC$.

\section{Classical simple Lie algebras}

Consider $\CC^n$ with coordinates $x^i$ and let $\d_i = \d/\d x^i$. The
algebra of general vector fields $\vect(n)$, generated by vector fields
$\xi = \xi^i(x)\d_i$, has the following infinite-dimensional simple
subalgebras:

\underline{$\vect(n)$}: 
Setting $\deg x^i = 1$ gives $\vect(n)$ a grading of depth $1$.
The grading operator $Z = x^i\d_i$ can be identified with dilatations.

\underline{$\svect(n)$}: 
Divergence-free vector fields which preserve the
volume form $V = dx^1 dx^2 ... dx^n = \eps_{ij..k} dx^i dx^j ... dx^k$.
Thus $\L_X V = 0$ for all $X \in \svect(n)$.
Setting $\deg x^i = 1$ gives $\svect(n)$ a grading of depth $1$.
The grading operator is $Z = x^i\d_i$.

\underline{$\hh(n)$, $n$ even}: 
Hamiltonian vector fields which preserve the symplectic two-form
$\w = \w_{ij} dx^i dx^j$, where the symplectic metric $\w_{ij} = -\w_{ji}$
and its inverse $\w^{ij}$ are structure constants. 
Thus $\L_X \w = 0$ for all $X \in \hh(n)$.
A Hamiltonian vector field is of the form $H_f = \w^{ij}\d_if\d_j$ 
where $f$ is a function, and the bracket in $\hh(n)$ reads explicitly
$[H_f, H_g] = H_{\{f,g\}}$, where the Poisson bracket reads
$\{f,g\} = \w^{ij}\d_if\d_jg$.
Setting $\deg x^i = 1$ gives $\hh(n)$ a grading of depth $1$.
The grading operator is $Z = x^i\d_i$.

\underline{$\kk(n+1)$, $n$ even}:
Denote the coordinates of $\CC^{n+1}$ by $t$ and $x^i$, $i=1,2,...,n$,
and let $\d_0$ and $\d_i$ denote the corresponding derivatives.
Contact vector fields preserve the space spanned by the contact 
one-form $\al = dt + \w_{ij} x^i dx^j$, where $\w_{ij}$ are the same
structure constants as in $\hh(n)$.
Thus $\L_X \al = f_X\al$, where $f_X$ is some function, for all 
$X \in \kk(n+1)$.
Setting $\deg x^i = 1$, $\deg t = 2$ gives $\kk(n+1)$ a grading of 
depth $2$.
The generators at non-positive degree are
\be
\barr{rclcl}
&\qquad&f&\qquad&K_f\\
\hline
\oj_{-2}: &&t&& H = \d_0, \\
\oj_{-1}: &&x_i&& P_i = \d_i - x_i\d_0, \\
\oj_0: && x_ix_j&&J_{ij} = J_{ji} = x_i\d_j + x_j\d_i, \\
&&t&&Z = 2t\d_0 + x^i\d_i,\\
\earr
\label{kfields}
\ee
where $x_i = \w_{ij}x^j$. 
$J_{ij}$ generate $\ssp(n)$ and $Z$ computes the grading.

Finite-dimensional simple algebras are subalgebras of $\ggl(n)$, which
in turn is the subalgebra of $\vect(n)$ that preserves the grading
operator $Z = x^i\d_i$, i.e. $X\in\ggl(n)$ iff $[Z,X] = 0$.

\underline{$\ssl(n)$}: 
$X\in\ssl(n)$ if $X\in\svect(n)$ and $[Z,X] = 0$.
$\dim\ssl(n) = n^2-1$.
$\rank\ssl(n) = n-1$.

\underline{$\ssp(n)$, $n$ even}: 
$X\in\ssp(n)$ if $X\in\hh(n)$ and $[Z,X] = 0$.
$\dim\ssp(n) = n(n+1)/2$.
$\rank\ssp(n) = n/2$.

\underline{$\sso(n)$}: 
The non-simple Poincar\'e algebra $\poin(n)$ consists of vector fields
which preserve the length element $ds^2 = g_{ij} dx^idx^j$, where
$g_{ij} = g_{ji}$ are symmetric structure constants. $\sso(n)$ is the
subalgebra at degree $0$, i.e.
$X\in\sso(n)$ if $X\in\poin(n)$ and $[Z,X] = 0$.
$\dim\sso(n) = n(n-1)/2$.
$\rank\sso(n) = n/2$ if $n$ even and 
$\rank\sso(n) = (n-1)/2$ if $n$ odd.

However, it is more interesting to describe these algebras together with
a non-trivial grading.

\underline{$\ssl(n+1)$}: 
This algebra admits a grading of depth $1$ of the form
$\ssl(n+1) = {\bf n^*} + (\ssl(n)+\ggl(1)) + {\bf n}$, where 
$\oj_{\pm1}$ are described as $\ssl(n)$ modules.
The generators are
\be
\barr{rcl}
\oj_{-1}: &\qquad& P_i = \d_i, \\
\oj_0: && J^i_j = x^i\d_j - {1\/n}\dlt^i_j x^k\d_k, \\
&&Z = x^i\d_i, \\
\oj_1: && K^i = x^i x^k\d_k. \\
\earr
\ee
The $\ssl(n+1)$ generators $J^\mu_\nu$ are identified as
\bes
J^0_0 = Z, &\qquad& J^i_j \to J^i_j - {1\/n}\dlt^i_jZ, \nle
J^0_i = P_i, && J^i_0 = -K^i.
\eens
$\dim\ssl(n+1) = n + n^2 + n = (n+1)^2-1$.
$\rank\ssl(n+1) = \rank\ssl(n)+\rank\ggl(1) = n$.

Note that $\ssl(n+1)$ contains all vector fields from $\vect(n)$ of
degree $0$ and $-1$, so non-trivial conditions only appear in $\oj_1$.
$X = X^i(x)\d_i \in\ssl(n+1)$ iff
\be
(n+1)\d_j\d_kX^i = \dlt^i_j\d_l\d_kX^l + \dlt^i_k\d_j\d_lX^l.
\ee
Since this is a second-order equation, $\ssl(n+1)$ does not preserve
some multi-linear form, which would yield a system of first-order 
equations. Phrased differently, it is a partial prolong rather than a
full prolong. Let $\Gamma^i_{jk}$ be (the inhomogeneous part of) a 
connection, which transforms under $\vect(n)$ as
\bes
\L_X \Gamma^i_{jk} &=& -X^l\d_l\Gamma^i_{jk} + \d_lX^i\Gamma^l_{jk} \nle
&&- \d_jX^l\Gamma^i_{lk} - \d_kX^l\Gamma^i_{jl} + \d_j\d_kX^i.
\eens
$\ssl(n+1)$ preserves the subspace 
spanned by the traceless part of $\Gamma^i_{jk}$,
\be
\gamma^i_{jk} = \Gamma^i_{jk} - 
{1\/n+1}(\dlt^i_j\Gamma^l_{lk} + \dlt^i_k\Gamma^l_{jl}).
\ee
Hence $X\in\ssl(n+1)\subset\vect(n)$ iff $\L_X \gamma^i_{jk}
= f^{i|mn}_{l|jk} \gamma^l_{mn}$ for some $X$-dependent functions
$f^{i|mn}_{l|jk}$.

\underline{$\co(n) = \sso(n+2)$}: 
The conformal algebra $\co(n)$ admits a grading of depth $1$ of the form
$\sso(n+2) = {\bf n^*} + (\sso(n)+\ggl(1)) + {\bf n}$, where 
$\oj_{\pm1}$ are described as $\sso(n)$ modules.
The generators are
\be
\barr{rcl}
\oj_{-1}: &\qquad& P_i = \d_i, \\
\oj_{0}: && J_{ij} = x_i\d_j - x_j\d_i,\\
&&Z = x^i\d_i, \\
\oj_{1}: && K_i = x_i x^k\d_k - \half x^kx_k\d_i, \\
\earr
\ee
where $x_i = g_{ij}x^j$.
$\co(n)$ preserves the space generated by the 
squared length element $ds^2 = g_{ij} dx^idx^j$, i.e. it
preserves the lightcone $ds^2 = 0$.
The $\sso(n+2)$ generators $J_{\mu\nu}$ are identified as
\bes
J_{0\bar0} = Z, &\qquad& 
J_{ij} = J_{ij}, \nle
J_{0i} = P_i &&
J_{\bar0 i} = -K_i.
\eens
$\dim\sso(n+2) = n + (n(n-1)/2+1) + n = (n+2)(n+1)/2$.
$\rank\sso(n+2) = \rank\sso(n)+\rank\ggl(1) = n/2+1$ if $n$ even.

\underline{$\ssp(n+2)$, $n$ even}:
This algebra admits a grading of depth $2$ of the form
$\ssp(n+2) = {\bf 1} + {\bf n} + (\ssp(n)+\ggl(1)) + {\bf n} + {\bf 1}$,
where $\oj_{\pm1}$ and $\oj_{\pm2}$ are described as $\ssp(n)$ modules.
More precisely, $X\in\ssp(n+2)$ if $X\in\kk(n+1)$ and 
\be
\L_X \beta^{ij} = f^{ij}_{kl}\beta^{kl},
\ee
for some $X$-dependent functions $f^{ij}_{kl}$, where 
\be
\beta^{ij} = dx^{[i}dx^{j]} + \w^{ij}\w_{kl}dx^k dx^l.
\label{sspbeta}
\ee
The generators are
\be
\barr{rcl}
\oj_{-2}: &\qquad& H = \d_0, \\
\oj_{-1}: && P_i = \d_i - x_i\d_0, \\
\oj_0: && J_{ij} = x_i\d_j + x_j\d_i, \\
&& Z = 2t\d_0 + x^i\d_i, \\
\oj_1: && K_i = x_i t\d_0 + x_ix^j\d_j - t\d_i, \\
\oj_2: && \Delta = t^2\d_0 + tx^i\d_i, \\
\earr
\label{spconf}
\ee
where $x_i = \w_{ij}x^j$. Note that the extra condition (\ref{sspbeta})
does not restrict the $\oj_{-2}+\oj_{-1}+\oj_0$ subalgebra, which is
the same as (\ref{kfields}).
One verifies that the generators in (\ref{spconf}) satisfy e.g.
\be
\barr{lclcl}
[P_i,P_j] = 2\w_{ij}H, &\qquad& [\Delta,P_i] = K_i, 
&\qquad& [H,\Delta] = Z, \\
{[}K_i,K_j] = -2\w_{ij}\Delta && [H,K_i] = -P_i,
&\qquad& [P_i, K_j] = J_{ij} - \w_{ij}Z.
\earr
\ee
The $\ssp(n+2)$ generators $J_{\mu\nu}$ are identified as
\be
\barr{lclcl}
J_{00} = H, &\qquad& J_{0i} = P_i, &\qquad& J_{0\bar0} = Z, \\
J_{\bar0\bar0} = \Delta, && J_{\bar0i} = K_i, && J_{ij} = J_{ij}.
\earr
\ee
$\dim\ssp(n+2) = 1 + n + (n(n+1)/2 +1) + n + 1 = (n+2)(n+3)/2$.
$\rank\ssp(n+2) = \rank\ssp(n)+\rank\ggl(1) = n/2+1$.

\section{General construction}
Consider $\CC^{n+1}$, $n$ even, with coordinates $t$ and $x^\mu$, where
$\mu = 1,2,...,n$. Denote the corresponding derivatives by 
$\d_0 = \d/\d t$ and $\d_\mu = \d/\d x^\mu$. A general vector field can
be written as $X = P^\mu\d_\mu + Q\d_0$ for some functions $P^\mu$ 
and $Q$.

Since $n$ is even we can introduce a non-degenerate constant 
symplectic metric $\Omega_{\mu\nu} = -\Omega_{\nu\mu}$. The contact 
algebra $\kk(n+1)$ is the subalgebra of $\vect(n+1)$ preserving the 
space  spanned by the contact form
\be
\alpha = dt + \Omega_{\mu\nu} x^\mu dx^\nu.
\label{alpha}
\ee
In other words, $\L_X\alpha = f_X\alpha$ for some functions $f_X$.
This condition leads to
\be
P^\mu = \Omega^{\mu\nu} \tD_\nu f, \qquad 
Q = 2f - x^\mu\tD_\mu f,
\ee
where $f$ is some function and
\be
\tD_\mu = \d_\mu + \Omega_{\mu\nu}x^\mu\d_0.
\ee
Alternatively, $\kk(n+1)$ can be defined as the subalgebra of 
$\vect(n+1)$ preserving the space spanned by $\tD_\mu$.

Now assume that the contact vector field $X$ also preserves the subspace
spanned by the bilinear form
\be
\beta^{ab} = \beta^{ab}_{\mu\nu} dx^\mu dx^\nu.
\label{quad}
\ee
Hence we require that $\L_X \beta^{ab} = g^{ab}_{cd}\beta^{cd}$ for some
$X$-dependent functions	$g^{ab}_{cd}$. One checks that this leads to 
the additional conditions
\bes
\beta^{ab}_{\mu\nu} \d_0 P^\mu &=& 0, \nle
\beta^{ab}_{\rho\nu} \d_\mu P^\rho + \beta^{ab}_{\mu\rho}\d_\nu P^\rho
&=& g^{ab}_{cd} \beta^{cd}_{\mu\nu}.
\eens
In terms of the function $f$, this becomes
\bes
\Omega^{\mu\rho}\beta^{ab}_{\mu\nu} \d_0\tD_\rho f &=& 0, \nle
\Omega^{\rho\si}\beta^{ab}_{\rho(\mu} \d_{\nu)}\tD_\si f &=& 
g^{ab}_{cd} \beta^{cd}_{\mu\nu}.
\eens

The question is now whether a non-trivial bilinear form (\ref{quad})
exists. It is clear that the structure constants $\Omega^{\mu\nu}$ and
$\beta^{ab}_{\mu\nu}$ must be built from natural constants only. If
$\oj_0$ can be described as $\ssl(n)$ plus some $\ssl(n)$ modules, the
only such constants are $\eps^{i_1i_2..i_n}$ and $\eps_{i_1i_2..i_n}$;
if $\oj_0$ includes $\ssp(n)$, we may also use the symplectic metric
$\w_{ij}$ and its inverse $\w^{ij}$. 

One bilinear form can always be constructed:
\be
\beta^{\mu\nu} = dx^{[\mu} dx^{\nu]} 
+ \Omega^{\mu\nu}\Omega_{\rho\si} dx^\rho dx^\si.
\label{sympbeta}
\ee
The algebra which preserves the spaces spanned by $\alpha$ and this
$\beta^{\mu\nu}$ can be identified with $\ssp(n+2)$, cf.
(\ref{sspbeta}). In the next five sections similar 
bilinear forms corresponding
to the five exceptional Lie algebras are described. Although I only
prove that bilinear forms yield the correct $\oj_{-2}+\oj_{-1}+\oj_0$
subalgebras, I have no doubt that the exceptions can be described in
this way; that it is possible to write down an exceptional bilinear 
form for each exceptional Lie algebra is highly non-trivial. Moreover, 
vector fields that preserve {\em some} structure {\em do} generate a 
subalgebra automatically.

One can also look for exceptional subalgebras of a conformal algebra
$\co(n) = so(n+2)$ instead of a contact algebra. Rather than preserving the
contact form (\ref{alpha}) up to a function, such vector fields preserve a
lightcone
\be
ds^2 \equiv \GG_{\mu\nu}dx^\mu dx^\nu = 0,
\ee
where $\GG_{\mu\nu} = \GG_{\nu\mu}$ is a symmetric metric. 
Such realizations, necessarily 
of depth $1$, are given for $\e_7$ and $\e_6$ in Sections \ref{e71} and
\ref{e61}. It is known that $\e_8$, $\f_4$ and $\g_2$ do not admit
gradings of depth $1$.

\section{$\e_8$, depth $2$}
\label{e8}
According to \cite{GKN00}, $\e_8$ admits a grading of depth $2$:
\bes
\e_8 &=& {\bf 1} + {\bf 56} + (\e_7 + \ggl(1)) + {\bf 56} + {\bf 1} \nle
&=&  {\bf 1} + ({\bf 28} + {\bf 28^*}) + (\ssl(8) + {\bf 70} + \ggl(1)) 
+ ({\bf 28} + {\bf 28^*}) + {\bf 1},
\eens
where the subspaces are described by their decomposition as $\e_7$ and
$\ssl(8)$ modules, respectively. Note that $\dim\e_7 = 133$ and 
$\dim\ssl(8) = 63$, so $\dim\e_8 = 248$.
$\rank\e_8 = \rank\e_7 + \rank\ggl(1) = 7+1 
= \rank\ssl(8)+ \rank\ggl(1) = 7+1 = 8$.

Consider $\CC^{57}$ with coordinates $t$, $x^{ij} = -x^{ji}$ and
$\bx_{ij} = -\bx_{ji}$, where $i,j=1,2,...,8$. Denote the corresponding
derivatives by $\d_0 = \d/\d t$, $\d_{ij} = \d/\d x^{ij}$ and
$\bd^{ij} = \d/\d\bx_{ij}$, where $\d_{kl}x^{ij} = \bd^{ij}\bx_{kl}
= \dlt^{ij}_{kl} \equiv \dlt^i_k\dlt^j_l - \dlt^j_k\dlt^i_l$.

The generators of $\oj_{-2} + \oj_{-1} + \oj_0$ are
\be
\barr{rcl}
\oj_{-2}: &\qquad& H = \d_0 \\
\oj_{-1}: && D_{ij} = \d_{ij} + \bx_{ij}\d_0, \\
&& E^{ij} = \bd^{ij} - x^{ij}\d_0, \\
\oj_{0}: && J^i_j = x^{ik}\d_{jk} - \bx_{jk}\bd^{ik}  
- \eighth \dlt^i_j (x^{kl}\d_{kl} - \bx_{kl}\bd^{kl}), \\
&& G^{ijkl} = x^{[ij}\bd^{kl]} - \eps^{ijklmnpq}\bx_{mn}\bd_{pq}, \\
&& Z = 2t\d_0 + \half x^{ij}\d_{ij} + \half \bx_{ij}\bd^{ij}.\\
\earr
\label{e8gen}
\ee
The subspace $\oj_1 + \oj_2$ is described in \cite{GKN00}; the 
explicit description is very tedious. Fortunately, it suffices to know
$\oj_{-2} + \oj_{-1} + \oj_0$ to determine the preserved structures.
Define 
\bes
\alpha &=& dt + \half x^{ij} d\bx_{ij} - \half \bx_{ij} dx^{ij}, 
\nlb{e8ab}
\beta^{ijkl} &=& dx^{[ij}dx^{kl]} + \eps^{ijklmnpq}d\bx_{mn}d\bx_{pq}.
\eens
$\alpha$ is the contact one-form and the bilinear form
$\beta^{ijkl} = \beta^{klij}$ is totally antisymmetric.
By direct calculation one shows 
that 
\bes
\L_X \alpha &=& f \alpha, 
\nlb{e8Lab}
\L_X \beta^{ijkl} &=& g^{ijkl}_{mnpq} \beta^{mnpq},
\eens
for some $X$-dependent functions $f$ and $g^{ijkl}_{mnpq}$. 
$\e_8$ can thus be described as the subalgebra of $\vect(57)$ which
preserves the spaces spanned by $\alpha$ and $\beta^{ijkl}$.

Moreover, the generators in (\ref{e8gen}) are the only vector fields of
degree $\leq 0$ with this property. To see this, note that contact 
algebra
$\kk(57)$ consists of vector fields that preserve the Pfaff equation
$\alpha=0$. All such vector fields can be written in the form
\be
X = K_f = \tE^{ij}f\tD_{ij} - \tD_{ij}f\tE^{ij} + 4fH,
\ee
where $f$ is a function and
\be
\tD_{ij} = \d_{ij} - \bx_{ij}\d_0, \qquad
\tE^{ij} = \bd^{ij} + x^{ij}\d_0.
\ee
In particular, vector fields of degree $\leq -1$ correspond to functions
of degree $\leq 1$, i.e.
\be
K_1 = 4H, \qquad K_{x^{ij}} = -2E^{ij}, \qquad K_{\bx_{ij}} = 2D_{ij}.
\ee
It remains to ensure that the vector fields of degree zero, corresponding
to the functions $x^{ij}x^{kl}$, $x^{ij}\bx_{kl}$, $\bx_{ij}\bx_{kl}$,
and $t$, are restricted to $J^i_j$, $G^{ijkl}$, and $Z$.  This is 
accompished by the second condition in (\ref{e8Lab}).

\section{$\e_7$, depth $2$}
\label{e72}
We obtain a realization of $\e_7$ from the previous section by restriction
$\ssl(8) \to \ssl(4)+\ssl(4)$:
\bes
\e_7 &=&  {\bf 1} + ({\bf 16} + {\bf 16^*}) 
+ (\ssl(4) + \ssl(4) + {\bf 36} + \ggl(1)) 
+ ({\bf 16} + {\bf 16^*}) + {\bf 1},
\eens
where the subspaces are described by their decomposition as 
$\ssl(4) + \ssl(4)$ modules; see \cite{GKN00}, Eq. (32).
Note that $\dim\ssl(4) = 15$, so $\dim\e_7 = 133$.
$\rank\e_7 = \rank\ssl(4)+ \rank\ssl(4) + \rank\ggl(1) = 3+3+1 = 7$.

Consider $\CC^{33}$ with coordinates $t$, $x^i_a$ and
$\bx^a_i$, where $i,j=1,2,3,4$ and $a,b=1,2,3,4$ are two different sets
of indices. Denote the corresponding
derivatives by $\d_0 = \d/\d t$, $\d^a_i = \d/\d x^i_a$ and
$\bd^i_a = \d/\d\bx^a_i$, where $\d_j^ax^i_b = \bd^i_b\bx^a_j
= \dlt^i_j\dlt^a_b$.

The generators of $\oj_{-2} + \oj_{-1} + \oj_0$ are
\be
\barr{rcl}
\oj_{-2}: &\qquad& H = \d_0 \\
\oj_{-1}: && D^a_i = \d^a_i + \bx^a_i\d_0, \\
&& E^i_a = \bd^i_a - x^i_a\d_0, \\
\oj_{0}: && I^i_j = x^i_a\d^a_j - \bx^a_j\bd^i_a 
-\fourth\dlt^i_j (x^k_a\d^a_k - \bx^a_k\bd^k_a), \\
&& J^a_b = \bx^a_i\bd^i_b - x^i_b\d^a_i
-\fourth\dlt^a_b (\bx^c_i\bd^i_c - x^i_c\d^c_i), \\
&& G^{ij}_{ab} = x^{[i}_{[a}\bd^{j]}_{b]} - 
\eps^{ijkl}\eps_{abcd}\bx^c_k\bx^d_l, \\
&& Z = 2t\d_0 + x^i_a\d^a_i + \bx^a_i\bd^i_a.\\
\earr
\label{e7gen}
\ee
Again, it suffices to know
$\oj_{-2} + \oj_{-1} + \oj_0$ to determine the preserved structures.
Define 
\bes
\alpha &=& dt + x^i_a d\bx^a_i - \bx^a_i dx^i_a, 
\nlb{e7ab}
\beta^{ij}_{ab} &=& dx^{[i}_{[a} dx^{j]}_{b]} + 
\eps^{ijkl}\eps_{abcd} d\bx^c_k d\bx^d_l.
\eens
$\alpha$ is the contact one-form and the bilinear
form $\beta^{ij}_{ab} = \beta^{ji}_{ba}$ is symmetric.
By direct calculation one shows that 
\bes
\L_X \alpha &=& f \alpha, 
\nlb{e7Lab}
\L_X \beta^{ij}_{ab} &=& g^{ij|cd}_{kl|ab} \beta^{kl}_{cd},
\eens
for some $X$-dependent functions $f$ and $g^{ij|cd}_{kl|ab}$. 
$\e_7$ can thus be described as the subalgebra of $\vect(33)$ which
preserves the spaces spanned by $\alpha$ and $\beta^{ij}_{ab}$.
The proof that (\ref{e7Lab}) uniquely singles out (\ref{e7gen}) among all
vector fields of degree $\leq0$ is completely analogous to the $\e_8$ 
case.

\section{$\e_6$, depth $2$}
According to \cite{GKN00}, Eq. (33), $\e_6$ admits a grading of depth $2$:
\bes
\e_6 &=& {\bf 1} + {\bf 20^*} + (\ssl(6) + \ggl(1)) 
+ {\bf 20} + {\bf 1},
\eens
where the subspaces are described by their decomposition as 
$\ssl(6)$ modules, respectively. Note that $\dim\ssl(6) = 35$,
so $\dim\e_6 = 78$.
$\rank\e_6 = \rank\ssl(6)+ \rank\ggl(1) = 5+1 = 6$.

Consider $\CC^{21}$ with coordinates $t$, $x^{ijk} = -x^{jik} = x^{jki}$,
where $i,j,k=1,2,...,6$. Denote the corresponding
derivatives by $\d_0 = \d/\d t$ and $\d_{ijk} = \d/\d x^{ijk}$, where 
$\d_{lmn}x^{ijk} = \dlt^{ijk}_{lmn} \equiv 
\dlt^{[i}_l\dlt^j_l\dlt^{k]}_m$.

The generators of $\oj_{-2} + \oj_{-1} + \oj_0$ are
\be
\barr{rcl}
\oj_{-2}: &\qquad& H = \d_0 \\
\oj_{-1}: && D_{ijk} = \d_{ijk} + \eps_{ijklmn}x^{lmn}\d_0, \\
\oj_{0}: && J^i_j = x^{ikl}\d_{jkl}
-\sixth \dlt^i_j x^{klm}\d_{klm}, \\
&& Z = 2t\d_0 + \sixth x^{ijk}\d_{ijk}.\\
\earr
\label{e6gen}
\ee
Let
\bes
\alpha &=& dt + \eps_{ijklmn} x^{ijk} dx^{lmn}, \nl
\beta_1^{ijk|lmn} &=& dx^{ijk}dx^{lmn} + dx^{lmn}dx^{ijk},
\nlb{e6ab}
\beta_2^{ijk|lmn} &=& \eps_{pqrstu}(
 \eps^{ijlmnq} dx^{krs} dx^{ntu} + [ijk] + [lmn]), \nl
\beta^{ijk|lmn} &=& \beta_1^{ijk|lmn} + \beta_2^{ijk|lmn},
\eens
where $[ijk]$ and $[lmn]$ stand for terms needed for proper
antisymmetrization.
The bilinear form $\beta^{ijk|lmn}$ has the symmetries
\be
\beta^{ijk|lmn} = -\beta^{jik|lmn} = \beta^{jki|lmn} = \beta^{lmn|ijk}.
\ee
One shows that 
\bes
\L_X \alpha &=& f \alpha, 
\nlb{e6Lab}
\L_X \beta^{ijk|lmn} &=& g^{ijk|lmn}_{pqr|stu} \beta^{pqr|stu},
\eens
for some $X$-dependent functions $f$ and $g^{ijk|lmn}_{pqr|stu}$. 
$\e_6$ can thus be described as the subalgebra of $\vect(21)$ which
preserves the spaces spanned by $\alpha$ and $\beta^{ijk|lmn}$.

\section{$\f_4$, depth $2$}
According to \cite{GKN00}, Eq. (34), $\f_4$ admits a grading of depth $2$:
\bes
\f_4 &=& {\bf 1} + {\bf 14} + (\ssp(6) + \ggl(1)) + {\bf 14} + {\bf 1},
\eens
where the subspaces are described by their decomposition as 
$\ssp(6)$ modules, respectively. 
Note that $\dim\ssp(6) = 20$, so $\dim\f_4 = 51$.
$\rank\f_4 = \rank\ssp(6)+ \rank\ggl(1) = 3+1 = 4$.
Since $\oj_0 = \ssp(6) + \ggl(1)$ preserves the symplectic two-form $\w$,
we can use the structure constants $\w_{ij} = -\w_{ji}$ to lower indices
and the inverse $\w^{ij}$ to raise them, e.g. $a_i = \w_{ij}a^i$,
$b^i = \w^{ij}b_j$.

Consider $\CC^{15}$ with coordinates $t$, $x^{ijk} = -x^{jik} =
x^{jki}$, where $i,j,k=1,2,...,6$ and $\w_{ij}x^{ijk} = 0$ for all $k$.
The number of $x$ indeterminates is thus $20-6 = 14$.
Denote the corresponding derivatives by $\d_0 = \d/\d t$ and 
$\d_{ijk} = \d/\d x^{ijk}$, where $\w^{ij}\d_{ijk} = 0$, 
$\d_{lmn}x^{ijk} = \dlt^{ijk}_{lmn} \equiv 
\dlt^{[i}_l\dlt^j_l\dlt^{k]}_m$.

The generators of $\oj_{-2} + \oj_{-1} + \oj_0$ are
\be
\barr{rcl}
\oj_{-2}: &\qquad& H = \d_0 \\
\oj_{-1}: && D_{ijk} = \d_{ijk} + \eps_{ijklmn}x^{lmn}\d_0, \\
\oj_{0}: && J_{ij} = \w_{ik}x^{klm}\d_{jlm} 
+ \w_{jk}x^{klm}\d_{ilm}, \\
&& Z = 2t\d_0 + \sixth x^{ijk}\d_{ijk}.\\
\earr
\label{f4gen}
\ee
Let
\bes
\alpha &=& dt + \eps_{ijklmn} x^{ijk} dx^{lmn}, \nl
\beta_1^{ijk|lmn} &=& dx^{ijk}dx^{lmn} + dx^{lmn}dx^{ijk},
\nlb{f4ab}
\beta_2^{ijk|lmn} &=& \eps_{pqrstu}(
 \eps^{ijlmnq} dx^{krs} dx^{ntu} + [ijk] + [lmn]), \nl
\beta^{ijk|lmn} &=& \beta_1^{ijk|lmn} + \beta_2^{ijk|lmn},
\eens
where $[ijk]$ and $[lmn]$ stand for terms needed for proper
antisymmetrization.
The bilinear form $\beta^{ijk|lmn}$ has the symmetries
\be
\beta^{ijk|lmn} = -\beta^{jik|lmn} = \beta^{jki|lmn} = \beta^{lmn|ijk}.
\ee
One shows that 
\bes
\L_X \alpha &=& f \alpha, 
\nlb{f4Lab}
\L_X \beta^{ijk|lmn} &=& g^{ijk|lmn}_{pqr|stu} \beta^{pqr|stu},
\eens
for some $X$-dependent functions $f$ and $g^{ijk|lmn}_{pqr|stu}$. 
$\f_4$ can thus be described as the subalgebra of $\vect(15)$ which
preserves the spaces spanned by $\alpha$ and $\beta^{ijk|lmn}$.

\section{$\g_2$, depth $2$}
According to \cite{GKN00}, Eq. (35), $\g_2$ admits a grading of depth $2$:
\bes
\g_2 &=& {\bf 1} + {\bf 4} + (\ssl(2) + \ggl(1)) + {\bf 4} + {\bf 1},
\eens
where the subspaces are described by their decomposition as 
$\ssl(2)$ modules, respectively. Note that $\dim\ssl(2) = 3$,
so $\dim\g_2 = 14$.
$\rank\g_2 = \rank\ssl(2)+ \rank\ggl(1) = 1+1 = 2$.

Consider $\CC^{5}$ with coordinates $t$, $x^{ijk} = x^{jik} = x^{jki}$,
where $i,j,k=1,2$. Denote the corresponding
derivatives by $\d_0 = \d/\d t$ and $\d_{ijk} = \d/\d x^{ijk}$, where 
$\d_{lmn}x^{ijk} = \dlt^{ijk}_{lmn} \equiv
\dlt^{(i}_l\dlt^j_m\dlt^{k)}_n$. 

Define the  structure constants
\bes
\Omega_{ijk|lmn} = \eps_{il}\eps_{jm}\eps_{kn} 
+ \eps_{jl}\eps_{km}\eps_{in} + \eps_{kl}\eps_{im}\eps_{jn}
\nle
+ \eps_{jl}\eps_{im}\eps_{kn} 
+ \eps_{kl}\eps_{jm}\eps_{in} + \eps_{il}\eps_{km}\eps_{jn},
\eens
which satisfy
\be
\Omega_{ijk|lmn} = \Omega_{jik|lmn} =\Omega_{jki|lmn} = -\Omega_{lmn|ijk}.
\ee

The generators of $\oj_{-2} + \oj_{-1} + \oj_0$ are
\be
\barr{rcl}
\oj_{-2}: &\qquad& H = \d_0 \\
\oj_{-1}: && D_{ijk} = \d_{ijk} + \Omega_{ijk|lmn}x^{lmn}\d_0, \\
\oj_{0}: && J^i_j = x^{ikl}\d_{jkl}
-\half \dlt^i_j x^{klm}\d_{klm}, \\
&& Z = 2t\d_0 + \sixth x^{ijk}\d_{ijk}.\\
\earr
\label{g2gen}
\ee
Let
\bes
\alpha &=& dt + \Omega_{ijk|lmn} x^{ijk} dx^{lmn}, \nl
\beta_1^{ij|kl} &=& \eps_{mn}dx^{ijm}dx^{kln}, \nl
\beta_2^{ij|kl} &=& \eps_{mp}\eps_{nq} (
\eps^{ik}dx^{jmn}dx^{lpq} + \eps^{jk}dx^{imn}dx^{lpq} +
\\
&&+ \eps^{il}dx^{jmn}dx^{kpq} + \eps^{jl}dx^{imn}dx^{kpq} ), \nl
\beta^{ij|kl} &=& \beta_1^{ij|kl} + \beta_2^{ij|kl}.
\eens
The bilinear form $\beta^{ij|kl}$ has the symmetries
\be
\beta^{ij|kl} = \beta^{ji|kl} = - \beta^{kl|ij}.
\ee
One shows that 
\bes
\L_X \alpha &=& f \alpha, 
\nlb{e4Lab}
\L_X \beta^{ij|kl} &=& g^{ij|kl}_{mn|pq} \beta^{mn|pq},
\eens
for some $X$-dependent functions $f$ and $g^{ij|kl}_{mn|pq}$. 
$\g_2$ can thus be described as the subalgebra of $\vect(5)$ which
preserves the spaces spanned by $\alpha$ and $\beta^{ij|kl}$.

\section{$\e_7$, depth $1$}
\label{e71}
According to \cite{GKN00}, Eq. (37), $\e_7$ also admits a realization of 
depth $1$:
\bes
\e_7 &=& {\bf 27} + (\e_6 + \ggl(1)) + {\bf 27} \nle
&=&  {\bf 27} + (\ssp(8) + {\bf 42} + \ggl(1)) + {\bf 27},
\eens
where the subspaces are described by their decomposition as $\e_6$ and
$\ssl(8)$ modules, respectively. Note that $\dim\e_6 = 78$ and 
$\dim\ssp(8) = 36$, so $\dim\e_7 = 133$.
$\rank\e_7 = \rank\e_6 + \rank\ggl(1) = 6+1 = 7$.
However, $\rank\ssp(8)+ \rank\ggl(1) = 4+1 = 5$. 

Since $\oj_0$ contains $\ssp(8)$ we have access to a symplectic metric
$\w_{ij} = -\w_{ji}$. It can be used to lower indices and its inverse 
$\w^{ij}$ raises them: $a_i = \w_{ij}a^i$ and $b^i = \w^{ij}b_j$.

An $\e_7$ realization is obtained from the $\e_8$ realization in 
Section \ref{e8} by imposing the conditions
\be
\bx_{ij} = \w_{ik}\w_{jl}x^{kl}, \qquad
\w_{ij} x^{ij} = 0.
\label{e71cond}
\ee
Upon this restriction, the contact form in (\ref{e8ab}) becomes
\be
\alpha = dt + \half x^{ij} \w_{ik}\w_{jl}dx^{kl}
- \half \w_{ik}\w_{jl}x^{kl} dx^{ij} = dt.
\ee
It is thus not possible to write down an interesting contact form.
However, with the structure constants at our disposal we can construct
the metric
\be
\GG_{ijkl} = \eps_{ijklmnpq}\w^{mn}\w^{pq},
\ee
which is totally antisymmetric, in particular
\be
\GG_{ijkl} = -\GG_{jikl} = \GG_{klij}.
\ee
We can thus consider vector fields that both preserve the lightcone
\be
ds^2 \equiv \GG_{ijkl} dx^{ij} dx^{kl} = 0
\ee
and the space spanned by $\beta^{ijkl}$ from (\ref{e8ab}). Since such
vector fields form a subalgebra of $\co(27)$, this realization is
of depth $1$.

The generators of $\oj_{-1} + \oj_0$ are
\be
\barr{rcl}
\oj_{-1}: && D_{ij} = \d_{ij}, \\
\oj_{0}: && J_{ij} = J_{ji} = 
\w_{ik}x^{kl}\d_{jl} + \w_{jk}x^{kl}\d_{il}, \\
&& G^{ijkl} = x^{[ij}\d^{kl]} - \eps^{ijklmnpq}x_{mn}\d_{pq}, \\
&& Z = \half x^{ij}\d_{ij}.\\
\earr
\label{e71gen}
\ee
Define 
\bes
ds^2 &=& \GG_{ijkl} dx^{ij} dx^{kl}, \nle
\beta^{ijkl} &=& dx^{[ij}dx^{kl]} + \eps^{ijklmnpq}dx_{mn}dx_{pq},
\eens
where $dx_{ij} = \w_{ik}\w_{jl}dx^{kl}$.
One shows that 
\bes
\L_X ds^2 &=& f ds^2, \nle
\L_X \beta^{ijkl} &=& g^{ijkl}_{mnpq} \beta^{mnpq},
\eens
for some $X$-dependent functions $f$ and $g^{ijkl}_{mnpq}$. 
$\e_7$ can thus be described as the subalgebra of $\vect(27)$ which
preserves the spaces spanned by $ds^2$ and $\beta^{ijkl}$.

\section{$\e_6$, depth $1$}
\label{e61}
The same idea as in the previous section also gives a restriction 
$\e_7 \to \e_6$ when applied to the $\e_7$ realization in 
Section \ref{e72}.
\bes
\e_6 &=& {\bf 16} + (\ssp(4) + \ssp(4) + {\bf 25} + \ggl(1)) + {\bf 16},
\eens
where the subspaces are described by their decomposition as 
$\ssp(4) + \ssp(4)$ modules.
Note that $\dim\ssp(4) = 10$, so $\dim\e_6 = 78$.
$\rank\e_6 = 6$ but 
$\rank\ssp(4)+ \rank\ssp(4)+ \rank\ggl(1) = 2+2+1 = 5$. 

Since $\oj_0$ contains $\ssp(4)+\ssp(4)$ we have access to two
symplectic metrics $\w_{ij} = -\w_{ji}$ and $\ww_{ab} = -\ww_{ba}$
and their inverses, which can be used to raise and lower indices:
$a_i = \w_{ik}a^k$, $b_a = \ww_{ab}b^b$,
$c^i = \w^{ik}c_k$, $d^a = \ww^{ab}d_b$.

This realization is obtained from the $\e_7$ realization by imposing
the condition
\be
\bx^a_i = \w_{ij}\ww^{ab}x^j_b.
\ee
Upon this restriction, the contact form in (\ref{e7ab}) becomes
\be
\alpha = dt + \w_{ij}\ww^{ab}x^i_a dx^j_b 
- \w_{ij}\ww^{ab}x^j_b dx^i_a = dt.
\ee
It is thus not possible to write down an interesting contact form.
However, with the structure constants at our disposal we can construct
the metric
\be
\GG^{ab}_{ij} = \eps_{ijkl}\eps^{abcd}\w^{kl}\ww_{cd},
\ee
which has the symmetries
\be
\GG^{ab}_{ij} = -\GG^{ab}_{ji} = -\GG^{ba}_{ij} 
= \GG^{ba}_{ji}.
\ee
We can thus consider vector fields that both preserve the lightcone
\be
ds^2 \equiv \GG^{ab}_{ij} dx^i_a dx^j_b = 0
\ee
and the space spanned by $\beta^{ij}_{ab}$ from (\ref{e7ab}). Since such
vector fields form a subalgebra of $\co(16)$, this realization is
of depth $1$.

The generators of $\oj_{-1} + \oj_0$ are
\be
\barr{rcl}
\oj_{-1}: && D^a_i = \d^a_i, \\
\oj_{0}: 
&& I_{ij} = I_{ji} = \w_{ik}x^k_a\d^a_j + \w_{jk}x^k_a\d^a_i, \\
&& J^{ab} = J^{ba} = \ww^{ac}x^i_c\d^b_i + \ww^{bc}x^i_c\d^a_i, \\
&& G^{ij}_{ab} = x^{[i}_{[a}\d^{j]}_{b]} - 
\eps^{ijkl}\eps_{abcd}x^c_k \d^d_l, \\
&& Z = x^i_a\d^a_i.\\
\earr
\ee
Define 
\bes
ds^2 &=& \GG^{ab}_{ij} dx^i_a dx^j_b,
\nle
\beta^{ij}_{ab} &=& dx^{[i}_{[a} dx^{j]}_{b]} + 
\eps^{ijkl}\eps_{abcd} dx^c_k dx^d_l,
\eens
where $dx^a_i = \w_{ij}\ww^{ab}dx^j_b$.
One shows that 
\bes
\L_X ds^2 &=& f ds^2, 
\nlb{e61Lab}
\L_X \beta^{ij}_{ab} &=& g^{ij|cd}_{kl|ab} \beta^{kl}_{cd},
\eens
for some $X$-dependent functions $f$ and $g^{ij|cd}_{kl|ab}$. 
$\e_6$ can thus be described as the subalgebra of $\vect(16)$ which
preserves the spaces spanned by $ds^2$ and $\beta^{ijkl}$.

\end{document}